\documentclass{ws-procs10x7}
\usepackage{balance}

\newcolumntype{d}[1]{D{.}{.}{#1}}

\def\Journal#1#2#3#4{{\it #1} {\bf #2}, #3 (#4)}

\makeindex
\begin{document}

\title{GLOBAL OBSERVABLES AT RHIC}

\author{A. MILOV}

\address{Physics Department, Brookhaven National Laboratory, Upton, NY 11973, USA\\E-mail: amilov@bnl.gov}


\twocolumn[\maketitle\abstract{Main
characteristics of the charged particle $dN_{ch}/d\eta$ and transverse
energy $dE_{T}/d\eta$ production measured in Heavy Ion collisions at
RHIC energies are presented in this article. Transformation of the
pseudo-rapidity shape, relation to the incident energy and centrality
profile are described in a systematic way. Centrality profile is
shown to be closely bound to the number of nucleons participating in the
collisions, at the same time an alternative approach to study the
centrality behavior is also discussed.}
]

\section{Introduction}
$dN_{ch}/d\eta$ and $dE_{T}/d\eta$ production are measured by all
RHIC experiments BRAHMS, PHENIX, PHOBOS and STAR. After several
years of RHIC operation data available from these experiments allow
to determine major dependencies in the global observables behavior in
High Energy Heavy Ion Collisions. This knowledge is essential for
the understanding of the physics of the Heavy Ions Collisions and for setting the
framework in which the study of the underlying Quark-Gluon Plasma
phase transition has to be carried out.

\section{The shape}
\begin{figure}[htb]
\centerline{\psfig{file=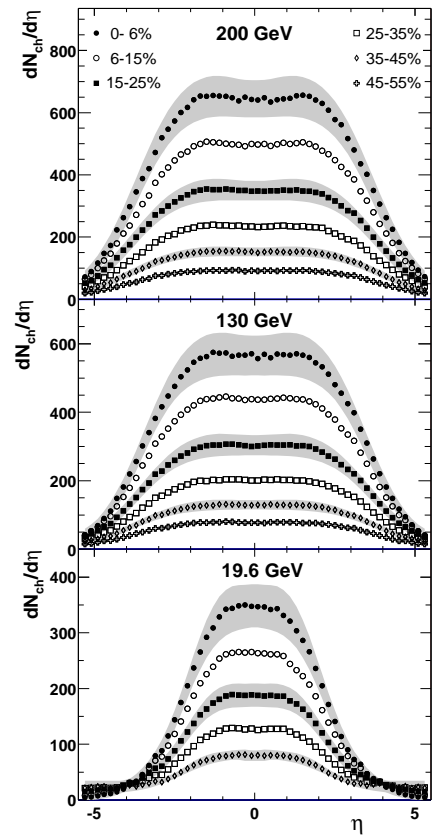,width=2.2in}}
\caption{$dN_{ch}/d\eta$ measured by PHOBOS in a full
range of pseudo-rapidity at several RHIC energies.}
\label{fig:wide-eta}
\end{figure}

The PHOBOS\cite{phobos-1} and BRAHMS\cite{brahms} experiments at RHIC measure the charged particle
distribution in a wide range of
pseudo-rapidity. Figure~\ref{fig:wide-eta} shows $dN_{ch}/d\eta$ for
different centralities in $Au+Au$ in the full range of
$\sqrt{s_{NN}}$ available at RHIC. 

All distribution presented in fig.~\ref{fig:wide-eta} have approximately a trapezoidal
shape with a characteristic parameter $y_{beam}$ equal to
$ln(\sqrt{s_{NN}}/m_{N})$, where $m_{N}$ is the mass of the
nucleon. One can define three different regions common to all
distributions shown in the figure. The flat top region including mid-rapidity
is between $\pm0.4y_{beam}$, the slope also referred as the
Limiting Fragmentation region around $\pm0.75y_{beam}$ and the
Fragmentation region close to $\pm 1.0y_{beam}$ and above.

At mid-rapidity where the most particles are produced the distribution increases
from peripheral to central collisions as one can see in all 3 panels in
fig.~\ref{fig:wide-eta}.  Opposite to that in the Fragmentation region
$dN_{ch}/d\eta$ decreases when collisions become more central as one can see it in the
$\sqrt{s_{NN}}=19.6$~$GeV$ (lower) panel\ of the figure.

\begin{figure}[htb]
\centerline{\psfig{file=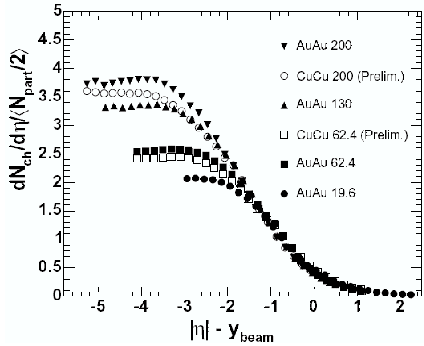,width=2.2in}}
\caption{Limiting Fragmentation Region measured for the same
centrality in different collision systems at different $\sqrt{s_{NN}}$.}
\label{fig:lfr}
\end{figure}

The Limiting Fragmentation Region located in between of the regions
with the opposite trends has a very distinct feature. In the rest
frame of the nuclei (in the $\eta-y_{beam}$ frame), the $dN_{ch}/d\eta$ is invariant to 
the incident energy and to the type of collision species for as long
as the centrality of the collisions are the same. This is demonstrated in fig.~\ref{fig:lfr}.
One shall note that this phenomenon extends on other collision systems\cite{peter1,ua1}
such as $p+p$ and $e+e$ where $\eta$ and $y_{beam}$ have to be redefined.

\section{Incident energy dependence}
\begin{figure}[htb]
\centerline{\psfig{file=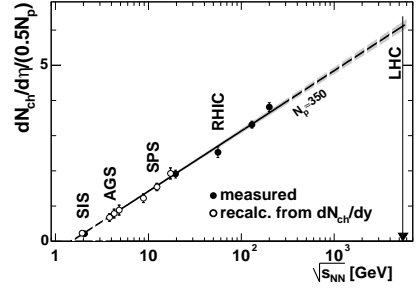,width=2.2in}}
\caption{$dN_{ch}/d\eta$ in the most central
collisions as a function of incident energy.}
\label{fig:log_s}
\end{figure}

As it is evident from the figs.~\ref{fig:wide-eta},\ref{fig:lfr}
particle production increases with the incident energy. At the beginning of
RHIC operation the PHENIX experiment suggested\cite{qm01} that this
rise in $dN_{ch}/d\eta$ is $\propto ln(\sqrt{s_{NN}}/m_{0})$ which
was later confirmed\cite{ppg19}. Dependence on the incident energy is shown in
fig.~\ref{fig:log_s}.

It is interesting to note the the logarithmic dependence holds for more
than 2 orders of magnitude in $\sqrt{s_{NN}}$ and work even for a
SIS energies\cite{ppg19,fopi} where only $\sim$100~$keV$ kinetic
energy is available per nucleon. Assuming the same behavior extends to the LHC energy
$\sqrt{s_{NN}}$=5.5~$TeV$ one would expect to observe $\sim1100$
charged particles in the most central events, equivalent to an
increase from the highest RHIC energy by $\sim$60\%.

\begin{figure}[htb]
\centerline{\psfig{file=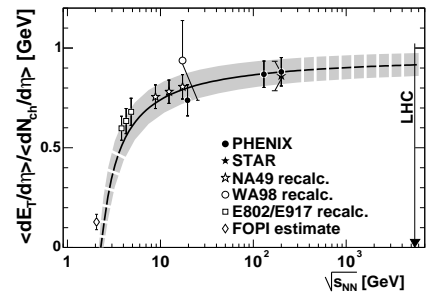,width=2.2in}}
\caption{$\langle dE_{T}/d\eta\rangle/\langle dN_{ch}/d\eta\rangle$
vs. $\sqrt{s_{NN}}$. Curve is the ratio of the fits discussed in the
text.}
\label{fig:et_nch_sqrts}
\end{figure}

The amount of the transverse energy produced in $Au+Au$
collisions\cite{ppg19} also logarithmically depends on
$\sqrt{s_{NN}}$. A small difference in $m_{0}$ fit
parameter,
$1.58\pm0.02$~$a.m.u.$ for $dN_{ch}/d\eta$ and $2.52\pm0.2$~$a.m.u.$ for $dE_{T}/d\eta$ 
produce a very distinct behavior of the ratio $\langle dE_{T}/d\eta
\rangle/\langle dN_{ch}/d\eta\rangle$ shown in fig.~\ref{fig:et_nch_sqrts}.
The ratio agrees well to the experimentally measured points and has
two different regimes. At low $\sqrt{s_{NN}}$ energy of produced
particle increases with the incident energy. Above $\sqrt{s_{NN}} \approx 10$~$GeV$
incident energy contributes to the number of produced particles,
while their energy grows much slower. If the same trend
preserves at LHC energies one would expect the ratio to increase only by
5\% compare to the highest RHIC energy. Since the $\langle dE_{T}/d\eta
\rangle/\langle dN_{ch}/d\eta\rangle$ is closely related to the freeze-out
temperature the LHC freeze-out conditions are expected to be similar
to RHIC.

\section{Centrality profile}
\begin{figure}[htb]
\centerline{\psfig{file=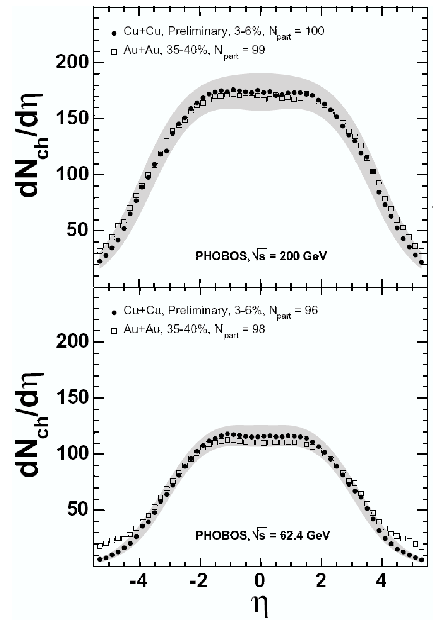,width=2.2in}}
\caption{$Cu+Cu$ and $Au+Au$ data at two RHIC energies selected in centrality bins with the same $N_{p}$.}
\label{fig:cucu-auau}
\end{figure}

The amount of matter available in a given collision mentioned in the
previous section is usually related to the number of participating
nucleons $N_{p}$. Since this relation is model dependent it was argued
since the beginning of RHIC operation whether $N_{p}$ is an adequate
parameter for the event characterization. The answer to this question can be found
by comparing PHOBOS data\cite{phobos-2} obtained in $Cu+Cu$ and in $Au+Au$
collisions shown in fig.~\ref{fig:cucu-auau}.
The $dN_{ch}/d\eta$ well agrees between $Cu+Cu$ and $Au+Au$ events
everywhere besides in the Fragmentation regions 
when one selects collisions with the same $N_{p}$. The
centrality bins correspond to the most central events in $Cu+Cu$ (impact
parameter $b\approx$2~$fm$) and to about $\sim$40\% central in
$Au+Au$ ($b\approx$9~$fm$).

\begin{figure}[htb]
\centerline{\psfig{file=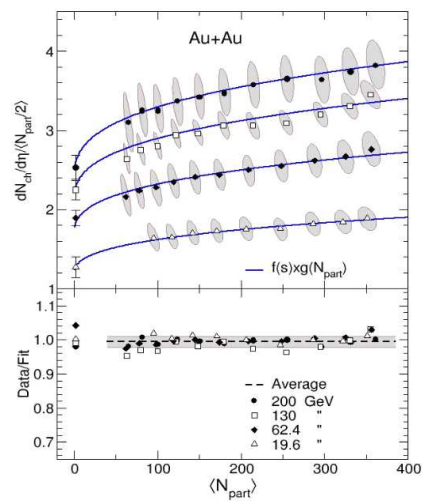,width=2.2in}}
\caption{Mid-rapidity $dN_{ch}/d\eta$ per pair of
$N_{p}$ at different RHIC energies measured by
PHOBOS and fitted with the same functional
form (top). Data divided by the fit (bottom).}
\label{fig:phobos-centr}
\end{figure}

Use of $N_{p}$ allows to study the centrality behavior of global
observables in a quantitative way. All four RHIC experiments provide
consistent\cite{ppg19} results for $dN_{ch}/d\eta$ centrality at
mid-rapidity as a function of $N_{p}$. PHOBOS experiment uses the same
functional form to fit its results measured at several RHIC
energies. This is shown in fig.~\ref{fig:phobos-centr}. 

\begin{figure}[htb]
\centerline{\psfig{file=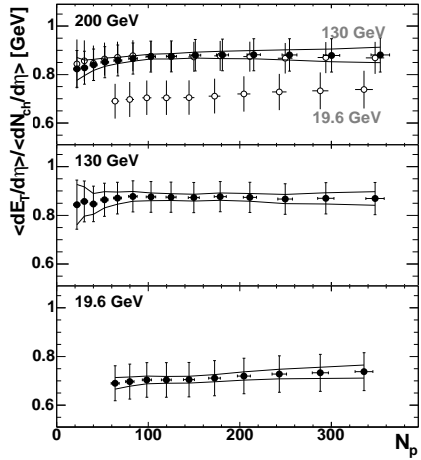,width=2.2in}}
\caption{ $\langle dE_{T}/d\eta\rangle/\langle
dN_{ch}/d\eta\rangle$ vs. $N_{p}$.}
\label{fig:et_nc_vs_npart}
\end{figure}

The magnitude of the curve is different at different $\sqrt{s_{NN}}$
according to the logarithmic trend as it was discussed above. The
shape of the curve as a function of $N_{p}$ remains the same at all
energies and well consistent with the data as one can see from the
data to fit ratio in the lower panel in the figure.
PHENIX\cite{ppg19} and STAR\cite{star} experiments measured centrality
dependence of the transverse energy production at RHIC. $dE_{T}/d\eta$ follows closely the shape of
the $dN_{ch}/d\eta$ such that the centrality profile of ratio $\langle dE_{T}/d\eta
\rangle/\langle dN_{ch}/d\eta\rangle$ is practically flat, see fig.~\ref{fig:et_nc_vs_npart}.

\begin{figure}[htb]
\centerline{\psfig{file=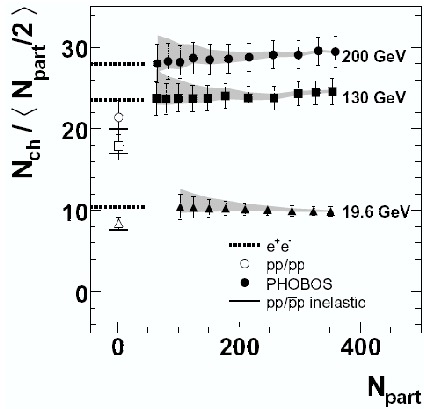,width=2.2in}}
\caption{Total $N_{ch}/N_{p}$ vs. $N_{p}$.}
\label{fig:nc_full}
\end{figure}

Total particle production over full pseudo-rapidity range
exhibits quite different centrality behavior than $dN_{ch}/d\eta$ at
mid-rapidity. PHOBOS data\cite{phobos-2} are shown in fig.~\ref{fig:nc_full}. For all measured
centralities it remains constant and $\sim$20\% higher than the $p+p$ (and $d+Au$) at the same
incident energy.
\vspace{-5mm}

\begin{figure}[htb]
\centerline{\psfig{file=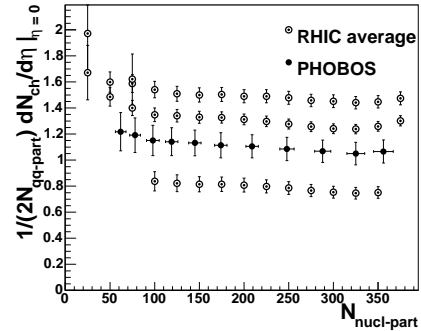,width=2.2in}}
\caption{Mid-rapidity $dN_{ch}/d\eta$ per pair of
participating quarks vs. number of participating nucleons.}
\label{fig:qq_part}
\end{figure}

In spite of a successful proof of the $N_{p}$ validity demonstrated in
fig.~\ref{fig:cucu-auau} it still remains a model dependent
quantity. In recent work\cite{voloshin} instead of the number of
nucleon participants, the number of quark participants is considered 
as a parameter to study the centrality dependence. The result is shown in fig.~\ref{fig:qq_part}.
As one can see instead of a rise of $dN_{ch}/d\eta$ per
nucleon participant shown in fig.~\ref{fig:phobos-centr}, the same per
quark participants remains flat over all measured centralities.

Work of the speaker is supported by the
Goldhaber Fellowship at BNL with funds provided by Brookhaven
Science Associates.

\end{document}